\documentstyle[prl,aps,epsf,rotate]{revtex}

\begin{document}
\twocolumn[\hsize\textwidth\columnwidth\hsize
     \csname @twocolumnfalse\endcsname

\title{On interference: the scalar problem}

\author{W. A. Hofer} 
\address{CMS-Vienna, Getreidemarkt 9/158, A--1060 Vienna, Austria \\
         {\bf email:} whofer@cms.tuwien.ac.at \quad {\bf web:}
           info.tuwien.ac.at/cms/wh/}

\maketitle


\begin{abstract} 
Single-slit and two-slit interferometer measurements of electrons are 
analyzed within the realistic model of particle propagation. In a step by
step procedure we show that all current models of interference are 
essentially non-local and demonstrate that the treatment of the quantum
theory of motion is the simplest model for the scalar problem. In particular
we give a novel interpretation of the quantum potential Q, which should be
regarded as a non-classical and essentially statistical term describing 
the changes of the quantum ensemble due to a change of the physical
environment. 
\end{abstract} 

\vspace{1 cm}

\vskip2pc]

\section{The nature of the problem}
\begin{figure}
\epsfxsize=0.9\hsize
\epsfbox{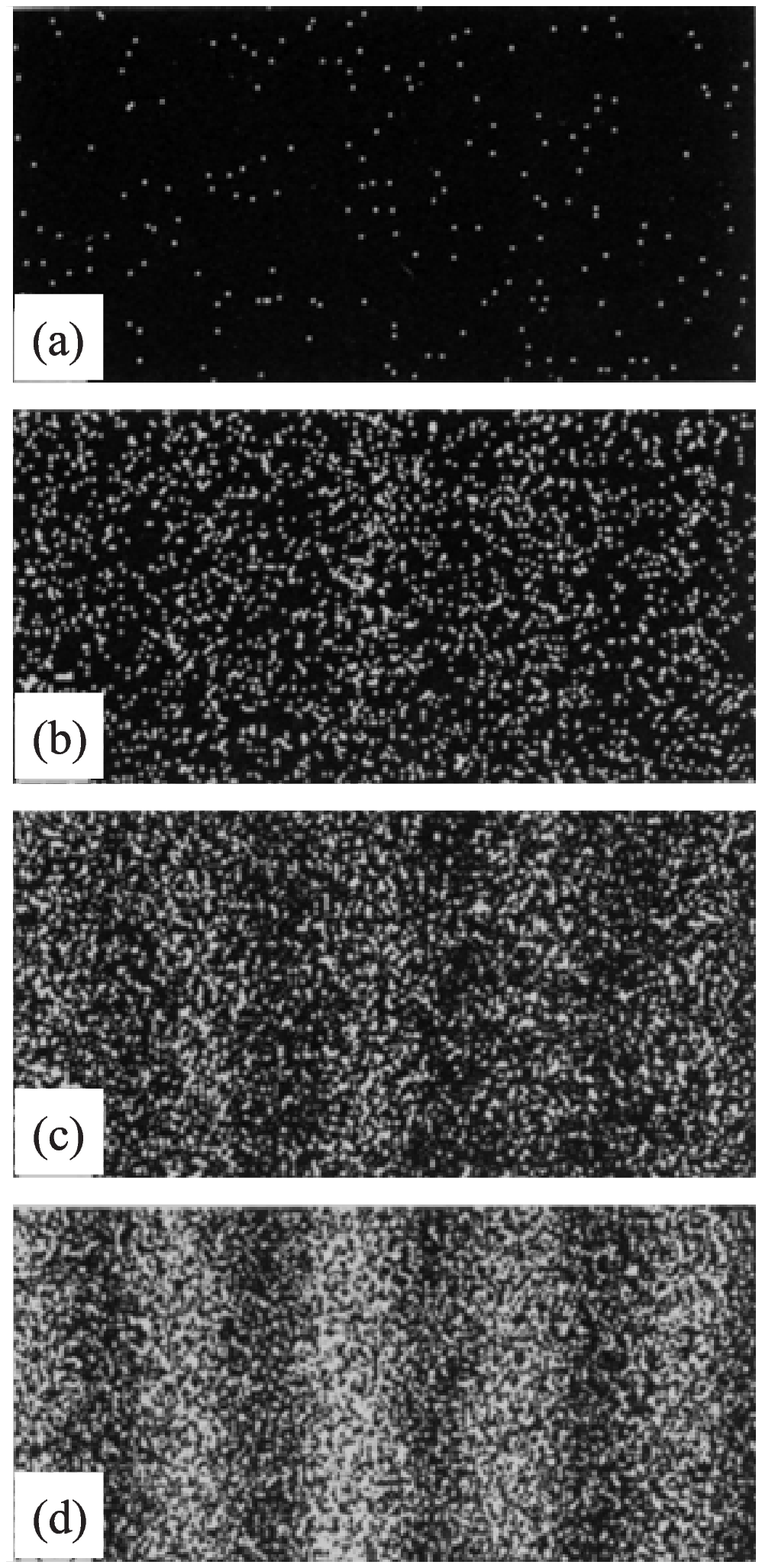}
\vspace{0.5cm}
\caption{Electron interference measurement according to Tonomura {\it et al.}
[1]. The cumulative pattern is generated by electrons sent one by
one through a two-slit interferometer. Number of electrons: (a) 100,
(b) 3000, (c) 20000, (d) 70000}
\label{fig001}
\end{figure}

Probably one of the most interesting problems in modern physics, since it
appears to be one of the most fundamental and, at the same time, one
of the simplest, is the problem of interference. The measurements of
Tonomura {\it et al.} \cite{tonomura89}, displayed in Fig. \ref{fig001},
reveal the impact of single
electrons on a detector screen, only gradually and without regularity evolving 
into the familiar interference pattern of a two-slit interferometer.
In Feynman's view this experiment ''has been designed to contain all of
the mystery of quantum mechanics, to put you up against the paradoxes
and mysteries and peculiarities of nature one hundred percent''
\cite{feynman63}. His judgement has gained, with these pictures,
a particularly striking illustration. The effect is commonly associated
with the notion of ''wave-particle duality'' \cite{selleri92} and is seen
as one of the  main examples, where our classical concepts of reality 
break down, since, after all, how can an electron passing through one
slit ''know'' whether the second aperture is opened,
a knowledge, it seems to possess, because its impact on the screen 
depends on the setup. These measurements can formally be described by
the quantum theory of motion \cite{philippidis79,philippidis82}, 
but although the
description is highly suggestive, it does not solve the problem, what the
''quantum-mechanical'' \cite{bohm52a} potential, substantial in this model,
is actually supposed to mean. In this sense the claim that the problem is
solved \cite{goldstein98} seems to be unjustified.

We propose, in this paper, a solution which is a step by step procedure,
where every step in the mathematical formalism can be justified by precise 
and conceptually consistent physical arguments. The solution will be based on
our recent analysis of measurement processes \cite{hofer99a}, which in turn is
founded on the notion of extended particles \cite{hofer98a}. It will be seen
that the solution given by the quantum theory of motion is formally correct,
although it requires a full understanding of the ensemble structure of 
quantum theory to be physically sound.

The paper is organized as follows: first we restate the classical solution to
a single slit experiment and show that the actual interference effect cannot 
be localized. Based on a conjecture about the interaction with the detector
screen we propose a measurement to detect an eventual incompleteness of
the standard model. Then the ensemble structure of quantum theory is briefly
discussed, and it will be shown that the probability interpretation of $\psi$
makes quantum theory (QT) inherently non-local. Based on the concept of quantum
ensembles a new interpretation will be given to the quantum potential Q
\cite{holland93}, and Q will be  found, as Bohm suggested \cite{bohm84},
related to information, since it describes the change of ensembles due to the
physical environment. Finally we shall analyze the solution for the 
two-slit interferometer in the quantum
theory of motion (QTM), in this case the analysis leads to the result that 
the simplest way to account for varying amplitudes and volumes of the 
extended and wave-like objects underneath the fundamental statements of QT
is to describe ensembles by varying densities of trajectories of point-like 
objects. In this sense the ''particle'' in QTM has a double meaning: it is
a single and well defined physical object but, due to the quantum potential,
also a member of the full quantum ensemble. It will be advocated that this
is the actual meaning of the term ''guiding-wave'', figuring prominently in
de Broglie's original concept \cite{broglie27}.

\section{Single-slit diffraction}
To analyze the main conceptual difficulties and 
the inherent non-locality in the classical
treatment of interference problems, it is sufficient to consider, in the
scalar approximation, a single slit experiment. We assume that the
electrons incident in our interferometer can be described as solutions
of the Helmholtz equation of their wave-function $\psi(\vec{r})$, the
wave shall cover a sufficient region for the relation to make sense.
With this definition we wish to avoid the discussion of boundaries and
coherence of the wave, since it will be seen that the fundamental
feature, which is the inherent non-locality, arises as soon as we 
admit a continuous wave-field as a suitable description. Helmholtz's
equation in the vacuum states:

\begin{equation}
\left(\Delta + k^2 \right) \psi(\vec{r}) = 0
\end{equation}

With Green's theorem and using the vacuum Green's function $\psi(\vec{r})$ is
transformed into an integral over the boundary R(V):

\begin{eqnarray}
\psi(\vec{r}) & = & \oint_{R(V)} d^2 \vec{f}' 
\frac{e^{ik |\vec{r} - \vec{r}'|}}{4 \pi |\vec{r} - \vec{r}'|}  \times
\nonumber \\
& \times & \left[ \nabla' \psi (\vec{r}') + 
\frac{2 \pi i \psi (\vec{r}')}{\lambda} 
\frac{\vec{r} - \vec{r}'}{|\vec{r} - \vec{r}'|}
\left( 1 + \frac{i}{k |\vec{r} - \vec{r}'|} \right) \right]
\end{eqnarray}

Using a $\delta$-functional to localize the wavelet in our system
we may rewrite $\psi(\vec{r})$ as:

\begin{equation}
\psi(\vec{r}') = \int_{- \infty}^{+ \infty} dt
\psi_{t}(\vec{r}',t) =:  \int_{- \infty}^{+ \infty} dt
\delta^3 \left(\vec{r}' - \vec{c}_{p}t\right) 
e^{ik |\vec{R} - \vec{c}_{p}t|}
\end{equation}

where - $\vec{R}$ is the source and $\vec{c}_{P}$ the velocity of the
''particle''. If we neglect the derivatives of $\delta$, which, after
all, shall only signify the existence of single entities, i.e. the fact that
a single electron has limited extension and is small compared to the system,
then we get at the moment $t = 0$:

\begin{eqnarray}
\psi_{t=0}(\vec{r}) & = & e^{ikR} \Gamma(k,\vec{r})  \\
\Gamma(k,\vec{r}) & = & \frac{i}{2 \lambda} \oint_{R(V)}
d^2 \vec{f}' \frac{\vec{r} - \vec{r}'}{|\vec{r} - \vec{r}'|^2} 
\left( 1 + \frac{i}{k |\vec{r} - \vec{r}'|} \right)
e^{ik |\vec{r} - \vec{r}'|} \nonumber
\end{eqnarray}

With a Kirchhoff approximation the calculation yields the familiar results
of classical electrodynamics: the amplitude $\Gamma(k,\vec{r})$ depends on the
setup geometry of the interferometer and the wavelength $ \lambda $. But it
provides the result already at the moment, when the ''particle'' passes the
slit (t = 0). Clearly, therefore, it does not describe the causal propagation
of single particles, but the probabilities of their impact on the screen, even if
this probability is contained in the intensity of the wave. A different way of
stating the same result is saying that the effect cannot be localized.

Now let the interferometer be operating in the Fraunhofer limit of diffraction
(distance between screen and slit sufficiently high), then the 
variations of the wavefunction are \cite{landau81}:

\begin{equation}
\psi (k,\theta) \propto \frac{\sin (k \theta)}{k \theta}
\end{equation}

where $k$ is the wavevector and $\theta$ the azimuthal angle.
A phase $\alpha$ of the wave at its origin $ - \vec{R}$ only adds a phasefactor,
which has no effect on the intensity in the conventional model. In the realistic
model \cite{hofer98a} it does have 
an effect, though, if the energy effecting the measurement is either only the
kinetic component or only the field component of particle energy.
Since the field component can be described as the real part
of the wave $ \psi$, the intensity from this energy component alone is given by:

\begin{equation}
dI (k, \theta, \alpha) = \left|
\frac{\sin (k \theta)}{k \theta} \right|^2
\cos^2 \alpha d \theta
\end{equation}

\begin{figure}
\epsfxsize=1.0\hsize
\epsfbox{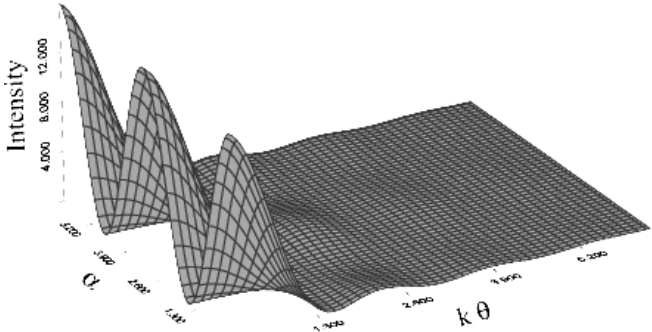}
\vspace{0.5cm}
\caption{Intensity of interference fringes due to the field component of
energy of the particle. The initial phase $\alpha$ determines the visibility
of the pattern on the screen in the far-field limit (Fraunhofer limit) of
the interferometer.}
\label{fig002}
\end{figure}

The intensity as a function of the azimuthal angle $\theta$ and the phase
$\alpha$ is shown in Fig. \ref{fig002}.
Therefore, a measurement of the interference fringes by using waves of a defined
phase may add to our information about the interaction process. Either the result
is independent of $ \alpha$, in this case the interaction energy can only be the 
total energy and the conventional model is sufficient, or the result depends,
like in Fig. \ref{fig002}, on the phase $ \alpha$, in this case the interaction
energy depends not on total energy but on an unbalanced composition of field
energy and kinetic energy. In either case the result yields an increase of 
information about the process: this applies to photons as well as electrons.

Summing up the result of this short classical analysis it can be said that 
in classical field theory the origin of the interference effect cannot be
localized, its result is obtained essentially by a summation of all components
over the whole system: it is therefore, in a sense, non-local, because the setup
can be chosen in such a way that the distance between the detector screen and the
slit environment is space-like for a given problem. And since the amplitude is
given already at the moment, when the photon passes the slit, its trajectory
contains information about the whole system in a non-local manner.

\section{Ensembles in quantum theory}

In non-relativistic QT a system is generally described by a suitable
Schr\"odinger equation \cite{schrodinger26}:

\begin{equation}
\hat{H} \psi = E \psi
\end{equation}

which we assume to be given also in case of a two-slit interferometer. It shall
consist of a Hamiltonian with potentials suitable to describe the slit environment.
The main conceptual problem here is that the single impacts, e.g. measured by
Tonomura {\it et al.} \cite{tonomura89}, are not described by the wavefunction
$ \psi (\vec{r})$, where $\vec{r}$ shall be a location on the detector screen,
since $\psi (\vec{r})$ only yields the overall intensity on the screen
but not the single hit. One does not need Einstein's fundamental criticism of
QT \cite{einstein35} to be left unsatisfied with a situation, where one 
observes events, which cannot be explained and even, if one follows the
orthodox \cite{bohr48}, are unexplainable in principle. 
Bohr, for example,
 considered any real event on the atomic level, which is, what these impacts
amount to, as beyond the means of scientific analysis, in his view ''such an
analysis is {\it in principle} excluded'' \cite{bohr49}.

We shall, in the following, interpret the single impacts as a result of the
ensemble structure of QT, which lies underneath its fundamental equations, and which
recently has been consistently analyzed for the first time \cite{hofer99a}.
The origin of the quantum ensemble is the unknown phase of $\psi$, together with
an intrinsic and field-like energy component of particle propagation
\cite{hofer98a}, not accounted for in the conventional framework of QT. In
particular it was shown that the uncertainty relations can be referred to
this intrinsic energy. Considering electrons of defined energy  where

\begin{equation}
E_{T} = \hbar \omega = m u^2
\end{equation}

is the total energy including the intrinsic components,
an external potential $V (\vec{r})$ leads to an ensemble wavefunction,
described by an integral over k-space:

\begin{eqnarray}
\psi (\vec{r}) & = & \frac{1}{(2 \pi)^{3/2}} \int_{0}^{k_{1}} d^3 k \, \chi_{0} (\vec{k})
e^{i \vec{k} \vec{r}} \nonumber \\
k_{1} & = & k_{1}(\vec{r}) = \sqrt{\frac{m}{\hbar^2} \left(E_{T} - V(\vec{r}) \right)}
\end{eqnarray}

The integral limit $k_{1}$ describes the error margin due to the undefined intrinsic
energy components in quantum theory.
Since the amplitude $\chi_{0} (\vec{k})$ is undefined, the condition that the density
of a single plane-wave component integrated over real space equals the mass of one
electron:

\begin{equation}
\int_{- \infty}^{+ \infty} d^3 r \left| \chi (\vec{r}, \vec{k}) \right|^2 = 
\chi_{0}^2 (\vec{k}) =: m_{e}
\end{equation}

leads, together with the probability interpretation of $\psi$ to the 
following integral:

\begin{eqnarray}
\int_{- \infty}^{+ \infty} d^3 r \left|\psi (\vec{r}, \vec{k}) \right|^2 & = &
\frac{m_{e}}{(2 \pi)^3} \int_{- \infty}^{+ \infty} d^3 r \times
\nonumber \\ & \times &
 \int_{0}^{k_{1}} d^3 k \, d^3k' e^{i \vec{r} (\vec{k}' - \vec{k})}
\end{eqnarray}

For a given potential in the system, $U = const.$ the integral reduces to
an integration over a sphere in k-space:

\begin{equation}
\int_{- \infty}^{+ \infty} d^3 r \left|\psi (\vec{r}, \vec{k}) \right|^2 =
\frac{4 \pi m_{e}}{3} k_{1}^3 \qquad
k_{1}  = \sqrt{\frac{m}{\hbar^2} \left(E_{T} - U\right)}
\end{equation}
 
In general the probability interpretation of $\psi$ thus has the following effects:
(i) The amplitude $\chi_{0} (\vec{k})$ has to be renormalized according to the
integration of $\psi$ over the whole system under the condition that 
\cite{born26}:

\begin{equation}
\int_{- \infty}^{+ \infty} d^3 r \left|\psi (\vec{r}, \vec{k}) \right|^2
=: 1
\end{equation}

And (ii) the ensemble wavefunction at a given location within the
system then depends on the potentials and amplitudes in all the other parts of
the system. A physical process, described via the wavefunction $\psi$ therefore
cannot be localized at a specific point $ \vec{r}$ of the system.

\section{Ensembles in the quantum theory of motion}

If we assume, as in QT, that the two-slit interference problem can be solved
by solving the Schr\"odinger equation:

\begin{equation}
\hat{H} \psi = E_{in} \psi
\end{equation}

where $\hat{H}$ shall be a suitable Hamiltonian and $\psi$ the wavefunction of
the problem, while $E_{in}$ denotes the kinetic energy of the incident electrons, then
the potentials of the problem should be known. Considering now, that $\psi$  is
the wavefunction of the ensemble, it carries two different informations: (i) The
information about the behavior of single electrons in the potential environment,
the {\em physical} side of the problem, and (ii) the information about the change
of the quantum ensemble due to the changes of the environment, the {\em statistical}
side of the problem.

Not considering the magnetic properties of electrons an analysis of Tonomura's
experiments may be limited to the scalar problem. Which means, that the polarizations
of intrinsic fields may be disregarded. From the viewpoint of single wavelets
the unknown properties are therefore the amplitude and/or the volume of single
''particles'' \cite{hofer99a,hofer98a}. For practical reasons we could assume
the volume of a single electron to be constant and, in the simplest possible
model, even point-like compared to all relevant distances within our system.
Then a single point-particle has to comply with two different constraints:
(i) The potentials in our system, the physical constraints, and (ii) the
ensemble structure of solutions of Schr\"odinger's equation due to the variation
of the ensemble with the external potentials. While the first is a strictly
classical term, in this formalization of the problem described by classical
mechanics, the latter is essentially non-classical and originates from the
probability interpretation of the wavefunction.
Comparing with the standard expressions of QTM, where the change of S,
the exponent of the ensemble wavefunction is described by:

\begin{equation}\label{eq015}
- \frac{\partial S}{\partial t} = \underbrace{V + \frac{(\nabla S)^2}{2 m_{e}} }_{I}
- \underbrace{ \frac{\hbar^2}{2 m_{e}} \frac{\nabla^2 R}{R} }_{II = Q}
\end{equation}

it is obvious that (I) describes the classical energy components of particle motion,
while (II), where R is the amplitude of the ensemble wavefunction:

\begin{equation}
\psi = R \cdot e^{i S/ \hbar}
\end{equation}

is a non-classical term. The main difference here is our interpretation of R and
Q. In QTM it is assumed that '' matter ... [is endowed] ... with a field aspect, mathematically
described by $R^2$ and $S$ ... which enables  ... [QTM] ... to avoid the paradox of an individual's
properties apparently depending on an ensemble." \cite{holland93_65}. In our view Q, the
''quantum mechanical'' \cite{bohm52a} potential is not due to some non-classical and still
physical field, but describes the change of the ensemble due to the change of the ''physical''
potential $V(\vec{r})$. In this sense it is a basically statistical term, even if it 
appears as an energy, and it is, as Bohm suspected \cite{bohm84}, related to the information
about the ensemble rather than any physical quality of the single ''particle''. The 
classical limit of motion then is the case, where $\psi$ does not influence the
trajectory $\vec{x}(t)$ of any single member.

\begin{figure}
\epsfxsize=1.0\hsize
\epsfbox{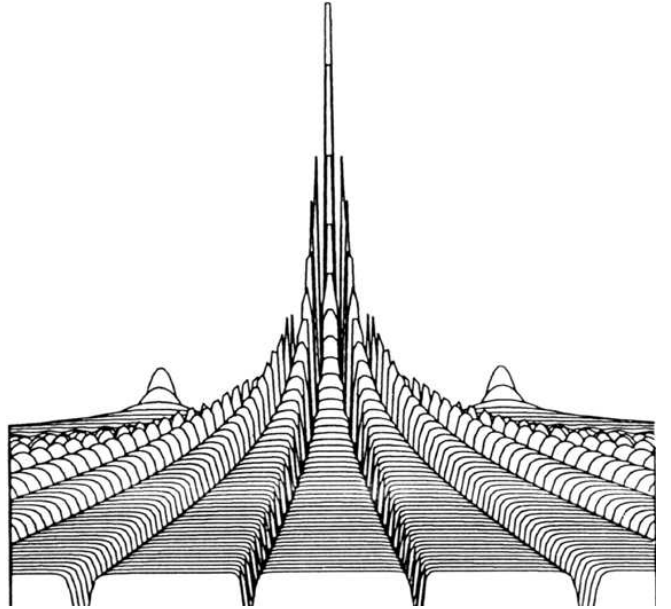}
\vspace{0.5cm}
\caption{Quantum potential for two Gaussian slits of the interferometer, the
potential is viewed from the detector screen, its origin is the change of the
quantum ensemble due to the change of the physical environment (from 
Philippidis {\it et al.} [4])}
\label{fig003}
\end{figure}

That this interpretation of Q is better suited to describe its real meaning than the
current one can also be seen from the fact that the ''particle equation of motion is
a deduction from the Schr\"odinger equation'' \cite{holland93_79}, but not vice versa, and,
most importantly, that although the particle '' responds to the local value of
the field in its vicinity (via Q) ... there is no reciprocal action of the particle
on the wave'' \cite{holland93_79}. If the quantum potential were of physical origin,
the last statement would be equivalent to a violation of Newton's third law 
({\it actio = reactio}). Therefore it cannot be of physical origin.

From a formal point of view the realistic interpretation and QTM consider the 
Schr\"odinger equation valid, and it has been held against Bohm's concept, that
it does not provide additional information \cite{zeh98}. In this case it must
be conceded that QTM is {\it logically} equivalent with standard QT. Which 
equally applies to the realistic interpretation. Then the result that the
Schr\"odinger equation is not an exact equation 
(a result of the realistic interpretation \cite{hofer98a}) 
requires the existence of 
additional terms describing the evolution of a system, and which are not related to
the evolution of any single - and well defined - member. And in this case one
arrives at the same conclusion: the additional term, showing up in QTM as
the quantum potential, can
{\it only} be related to this evolution of the system, described by the 
Schr\"odinger equation. In this sense it is also correct to say that QT
does not know any well defined object, because of its fundamental equation.

\begin{figure}
\epsfxsize=1.0\hsize
\epsfbox{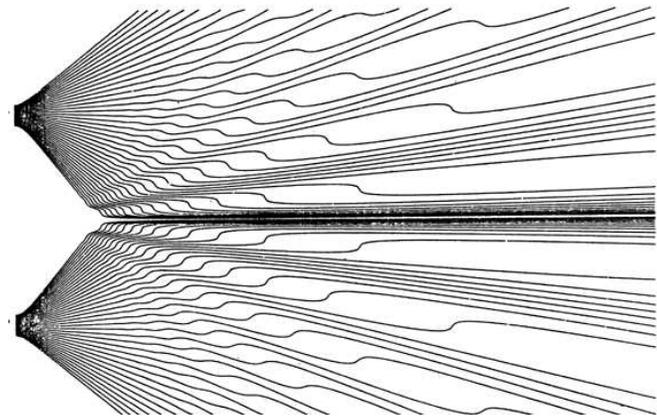}
\vspace{0.5cm}
\caption{Trajectories of single particles in a two-slit
interferometer from [4]. The unusual curvature of the
trajectories is due to the quantum potential Q.}
\label{fig004}
\end{figure}

It is interesting to note that, as Peter Holland pointed out, most orthodox
physicists attack the quantum theory of motion for two, mutually exclusive,
reasons: for being too classical (in its conception of particles, trajectories
etc.) and for being not classical enough (in its non-locality, the concept of
the quantum potential etc.) \cite{holland93_26}. Zeh recently gave a similar
example in an article with the programmatic title: ''Why Bohm's Quantum Theory?''
\cite{zeh98}, when he criticized the theory for being logically equivalent
to QT (''successful only because it keeps Schr\"odinger's (exact) wave mechanics'')
and going beyond QT (although ''the rest of it is observationally meaningless'',
the trajectories it describes are thought to result from ''unobservable causes
for stochastic events''). Now QTM is, of course, logically equivalent to the
Schr\"odinger equation. And once it is accepted, that it is a simplified account
of events - the point particle is the simplest possible model for the single
wavelets in QT -, then it teaches us something new, since it combines
the notion of an individual with the notion of an ensemble in a consistent
and instructive manner - something QT has not been able to achieve in seventy
years of arguing. The GRW model \cite{ghirardi86} mentioned 
can not be considered on an equal footing, since it depends on a
non-linear addition to the Hamiltonian: where this term should
actually come from, remains a mystery, even if Zeh refers it to
''fundamentally irreversible {\it spontaneous localization}''.

In his treatment of the two-slit problem Philippidis begins by calculating the
wavefunction $ \psi$ after the slits, assumed Gaussian for convenience, 
and which is then  employed to derive the quantum potential Q from Eq. (\ref{eq015}).
The numerical calculation used characteristic experimental data
of a 1961 electron interference measurement by J\"onsson \cite{joensson61}, 
it was calculated for the whole region between the slits and the detector screen.
It is essentially due to the local constraints of the wavefunction at the two-slits,
contrary to the previous chapter, where our main emphasis was on constraints due to
external potentials. Given the symmetry of QT (and also QTM) concerning real space
and momentum space, the difference should not change the picture: a change of the
ensemble gives rise, in QTM, to a quantum potential Q. 
Fig. \ref{fig003} reproduces the result of the calculation, the two parabolic
peaks in the back coincide with the slit positions. The particle trajectories in
QTM are calculated by integrating the equation:

\begin{equation}
\vec{p} = \nabla S
\end{equation}

where $\vec{p}$ is the momentum of the particle and $S$ the exponent of the wavefunction.
Initially, the trajectories from the two slits fan out like for a single slit
interference measurement, it is only where Q becomes appreciable, that distinct 
kinks appear, which are due to a rapid acceleration at the troughs of the quantum
potential. The trajectories of single particles are displayed in Fig.
\ref{fig004}. The single hits on the screen reproduce the overall pattern of the intensity
calculated in the conventional manner, but the single impacts are at distinct
locations: in that sense QTM reproduces the full extent of the experimental findings
(in contrast to QT, where only the intensity is given). And while in QT the electron
is a particle and a wave simultaneously, it is in QTM a particle, guided by its 
quantum potential Q, which represents the full quantum ensemble of a given
environment. 

\section{Non-locality, or what?}

The question, whether or not an individual electron ''knows'' of the
setup, most remain open, since, as the analysis reveals, none of our
current theories gives a local description of the problem. In classical
field theories the fact has been known for quite some time, this is,
what we mean, when we speak about ''waves'', but in QT, due to the
essentially abstract framework, the same feature is less obvious. 

Non-locality, it has been shown, enters the framework essentially via
the normalization of the wavefunction, because this cannot be done without
considering the ensemble over the whole system. 
But while in a classical context the non-locality could be argued on the
basis of the rather high field extension, this seems no longer a possibility
considering current experimental techniques. With femto-lasers and,
most strikingly, atom interference measurements \cite{leibfried98} the
typical extension of a single wavelet is well below the separation of 
the slits of an interferometer. Conceptually, one either has to concede
that the formal non-locality is also an experimental fact, because it is
hard to support the notion that interactions of an atom with one slit-environment
depend on the arrangement of atoms some $10^4$ atom diameters away
(distance of the second slit about 8 $\mu m$ \cite{leibfried98}). Or,
along a completely different line of reasoning, one argues that the
coherence of the beam is the ultimate origin of non-locality, since it
guarantees that a single atom fits into the overall pattern (e.g. via its phase).
A possibility, which we shall explore in future publications.

The main improvement, compared to the current state of affairs in QTM,
is conceptual. While, for example, the calculation of the quantum potential
(a {\it physical} cause for the motion of a single particle) from the
wavefunction (according to current belief \cite{born26} a {\it statistical}
measure of events and locations) must remain logically inconsistent,
the procedure becomes perfectly sound, once the wavefunction gains a double
meaning (a fundamental result of the realistic interpretation \cite{hofer99a})
and if we concede that the point-particle is only a very crude approximation.
In this sense Bohm's quantum theory of motion appears to be the simplest
mathematical form the realistic interpretation can take. Which means also
that it's extension to all fields, e.g. to the atomic domain, might not make
too much sense. Because in fundamental processes, where the physics of a system
are as well known as in hydrogen \cite{hofer98b}, the simplifications
of the quantum theory of motion may well
lead to a completely distorted picture.




\begin{references}
\bibitem{tonomura89}
Tonomura A., Endo J., Matsuda T., Kawasaki T., and Ezawa H.
{\it Am. J. Phys.} {\bf 57}, 117 (1989)
\bibitem{feynman63}
Feynman R. P., Leighton R. B., Sands M.
{\it The Feynman Lectures on Physics}, vol. 1, Addison-Wesley,
New York (1963) p.37
\bibitem{selleri92}
Selleri F. (ed) {\it Wave-particle duality}, Plenum Press,
New York (1992)
\bibitem{philippidis79}
Philippidis C., Dewdney C., and Hiley B. J.
{\it Nuovo Cimento} {\bf 52B}, 15 (1979)
\bibitem{philippidis82}
Philippidis C., Bohm D., and Kaye R. D.
{\it Nuovo Cimento} {\bf 71B}, 75 (1982)
\bibitem{bohm52a}
Bohm D. {\it Phys. Rev.} {\bf 85}, 166 (1952)
\bibitem{goldstein98}
Goldstein S. {\it Physics Today} 42 (March) and 38 (April) (1998)
\bibitem{hofer99a}
Hofer W. A. {\it Measurements in quantum physics: the interplay
between physical properties and statistics}, submitted (1999),
preprint at quant-ph/9704006
\bibitem{hofer98a}
Hofer W. A. {\it Physica A} {\bf 256}, 178 (1998)
\bibitem{holland93}
Holland P. R. {\it The quantum theory of motion},
Cambridge University Press, Cambridge UK (1993)
\bibitem{bohm84}
Bohm D. and Hiley B. J. {\it Found. Phys.} {\bf 14}, 255 (1984)
\bibitem{broglie27}
De Broglie L. {\it J. Phys.} {\bf 8}, 225 (1927)
\bibitem{landau81}
Landau L. D. and Lifschitz E. M.
{\it Klassische Feldtheorie}, Springer Berlin (1981) p. 185
\bibitem{schrodinger26}
Schr\"odinger E. {\it Ann. Physik} {\bf 79}, 361 and 489 (1926)
\bibitem{einstein35}
Einstein A., Rosen N., and Podolsky B. {\it Phys. Rev.}
{\bf 47}, 180 (1935)
\bibitem{bohr48}
In {\it Dialectica} {\bf 2}, 312 (1948) Niels Bohr formulated his
point of view as follows:''The entire formalism is to be considered as a
tool for deriving predictions ... the symbols themselves, as is
indicated already by the use of imaginary numbers, are not susceptible
to pictorial interpretation''.
\bibitem{bohr49}
Bohr N. in Schilpp P. A.  (ed) {\it Albert Einstein,
Philosopher-Scientist}, Evanston, Ill. (1949)
\bibitem{born26}
Born M. {\it Z. Physik} {\bf 37}, 863 (1926)
\bibitem{holland93_65}
Holland P. R. (1993) p. 65
\bibitem{holland93_79}
Holland P. R. (1993) p. 79
\bibitem{zeh98}
Zeh H. D. {\it Found. Phys. Lett.} {\bf 12}, 197 (1999) 
\bibitem{holland93_26}
Holland P. R. (1993) p. 26
\bibitem{ghirardi86}
Ghirardi G.C., Rimini A., and Weber D. {\it Phys. Rev. D} {\bf 34}, 470 (1986)
\bibitem{joensson61}
J\"onsson C. {\it Z. Physik} {\bf 161}, 454 (1961)
\bibitem{leibfried98}
Leibfried D., Pfau T., and Monroe C. {\it Physics Today}, 22 (April) (1998)
\bibitem{hofer98b}
Hofer W. A. {\it A dynamic model of atoms: structure, internal interactions
and photon emissions of hydrogen}, preprint at quant-ph/9801044
\end{references}
\end{document}